\documentclass[12pt]{article}

\usepackage{amsmath}
\usepackage{amsfonts}
\usepackage{amssymb}
\usepackage{latexsym}
\usepackage{enumerate}
\usepackage{graphicx}
\usepackage[english]{babel}

\newcommand{\captionfonts}{\small}

\makeatletter  
\long\def\@makecaption#1#2{%
  \vskip\abovecaptionskip
  \sbox\@tempboxa{{\captionfonts #1: #2}}%
 \ifdim \wd\@tempboxa >\hsize
    {\captionfonts #1: #2\par}
  \else
    \hbox to\hsize{\hfil\box\@tempboxa\hfil}%
  \fi
  \vskip\belowcaptionskip}
\makeatother   

\usepackage[
      colorlinks=true,
      linkcolor=black,
      urlcolor=blue,
      filecolor=blue,
      citecolor=blue,
      pdfstartview=FitH,
      pdftitle={},
      pdfauthor={},
      pdfsubject={},
      pdfkeywords={},
      pdfpagemode=UseNone,
      bookmarksopen=true
      ]{hyperref}
\usepackage[all]{hypcap}     

\begin{document}
\input epsf

\def\p{\partial}
\def\h{{1\over 2}}
\def\be{\begin{equation}}
\def\bea{\begin{eqnarray}}
\def\ee{\end{equation}}
\def\eea{\end{eqnarray}}
\def\d{\partial}
\def\la{\lambda}
\def\eps{\epsilon}
\def\bb{\bigskip}
\def\mm{\medskip}
\newcommand{\dm}{\begin{displaymath}}
\newcommand{\edm}{\end{displaymath}}
\renewcommand{\b}{\tilde{B}}
\newcommand{\gm}{\Gamma}
\newcommand{\ac}[2]{\ensuremath{\{ #1, #2 \}}}
\renewcommand{\ell}{l}
\newcommand{\z}{\ell}
\newcommand{\newsection}[1]{\section{#1} \setcounter{equation}{0}}
\def\bb{$\bullet$}
\def\Qbar{{\bar Q}_1}
\def\QPbar{{\bar Q}_p}

\def\q{\quad}

\def\bn{B_\circ}

\let\a=\alpha \let\b=\beta \let\g=\gamma \let\d=\delta \let\e=\epsilon
\let\c=\chi \let\th=\theta  \let\k=\kappa
\let\l=\lambda \let\m=\mu \let\n=\nu \let\x=\xi \let\r=\rho
\let\s=\sigma \let\t=\tau
\let\vp=\varphi \let\vep=\varepsilon
\let\w=\omega      \let\G=\Gamma \let\D=\Delta \let\Th=\Theta
                     \let\P=\Pi \let\S=\Sigma

\def\h{{1\over 2}}
\def\t{\tilde}
\def\r{\rightarrow}
\def\nn{\nonumber\\}
\let\bm=\bibitem
\def\Kt{{\tilde K}}
\def\b{\bigskip}

\let\p=\partial

\begin{flushright}
\end{flushright}
\vspace{20mm}
\begin{center}
{\LARGE  Comments on black holes I:}
\\
\vspace{5mm}
{\LARGE  The possibility of complementarity}
\\
\vspace{18mm}
{\bf  Samir D. Mathur  ~and~ David Turton }\\

\vspace{8mm}
Department of Physics,\\ The Ohio State University,\\ Columbus,
OH 43210, USA\\ \vskip .2 in   mathur.16@osu.edu\\turton.7@osu.edu
\vspace{4mm}
\end{center}
\vspace{10mm}
\thispagestyle{empty}
\begin{abstract}
\b
We comment on a recent paper of Almheiri, Marolf, Polchinski and Sully  who argue against black hole complementarity based on the claim that an infalling observer `burns' as he attempts to cross the horizon.  We show that measurements made by an infalling observer outside the horizon are statistically identical  for the cases of  vacuum  at the horizon and radiation emerging from a stretched horizon. This forces us to follow the dynamics all the way to the horizon, where we need to know the details of Planck-scale physics. We note that in string theory the fuzzball structure of microstates does not give any place to `continue through' this Planck regime.  AMPS argue that interactions near the horizon preclude traditional complementarity. But the conjecture of `fuzzball complementarity' works in the opposite way: the infalling quantum is absorbed by the fuzzball surface, and it is the resulting dynamics that is conjectured to admit a complementary description.

\b

\end{abstract}
\vskip 1.0 true in

\newpage

\numberwithin{equation}{section}
\setcounter{tocdepth}{1}
\tableofcontents

\baselineskip=16pt
\parskip=3pt

\section{Introduction}
 
 The quantum theory of black holes has proven to be rich territory for the exploration of the most fundamental laws of physics. The discoveries of black hole entropy \cite{bek}, and  Hawking radiation \cite{hawking} provided deep links between gravity and thermodynamics, while raising a serious problem in the form of the information paradox. One suggestion that arose in this context was the notion of black hole complementarity \cite{complementarity}. String theory provides a microscopic explanation for the entropy of black holes \cite{sv}, and the fuzzball structure of microstates provides a solution to the information paradox \cite{lm4,fuzzballs,fuzzball3,fuzzball4,cern,plumpre,plumberg,otherfcrefs}.
 
 Recently there have appeared several papers  discussing the relations between the information paradox, entanglement theorems, complementarity and other issues involving the quantum theories of black holes \cite{amps,ampsfollowups}\footnote{See also the earlier work of \cite{Braunstein:2009my}.}. Since there are several interrelated issues in the area of black holes, we have split our discussion into a set of papers, each addressing a different question. In this article we comment on some of the arguments used in the paper of Almheiri, Marolf, Polchinski and Sully (AMPS) \cite{amps} and argue that they do not address the conjecture of `fuzzball complementarity' developed in \cite{plumpre,plumberg,otherfcrefs}.
 
 We note that the fuzzball program provides a  consistent picture of all issues in the quantum dynamics of black holes (see \cite{reviews} for reviews). We will keep this fact at the back of our mind, since in many cases the fuzzball description provides us an explicit model to judge the validity of abstract arguments. 
  
 We begin with some definitions and basic facts about black holes and the information paradox. We then make two observations:

 (a) It is often assumed that if an infalling observer `hits something' at the horizon, then there cannot be a `complementary' description where he goes through. While traditional complementarity may have this feature, the kind of complementarity  suggested by fuzzballs is different. We use a toy example provided by  AdS/CFT duality to observe that in one description an infalling quantum `breaks up', while in another description it continues its trajectory unscathed.  We note that the case of the black hole is somewhat different from the AdS/CFT case, and explain how complementarity can arise for hard-impact processes involving quanta with energy $E\gg kT$ falling freely into the black hole\footnote{Here $E$ refers to the conserved Killing energy of the infalling quantum, and $T$ is the temperature of the black hole as measured from infinity.}.

 (b) One might think that an observer falling into the traditional black hole sees nothing as he falls up to the horizon, but an observer falling towards a body radiating `real quanta' from a stretched horizon  would get `burnt' by the highly energetic photons encountered close to this horizon. We show that observations of Hawking quanta made outside the horizon actually yield  similar results in both cases. Switching off a detector before crossing the horizon of a traditional black hole creates excitations from vacuum fluctuations, and these excitations have the same spectrum as excitations created by `real quanta' from a stretched horizon.

 We then address the argument made in  AMPS \cite{amps}. In  brief outline, the AMPS argument goes as follows: 
\begin{enumerate}[(i)]
	\item If Hawking evaporation is unitary, then the state near the horizon is not the vacuum in an infalling observer's frame, but involves high-energy excitations.
	\vspace{-3mm}
	\item If there are high-energy excitations near the horizon, then an  infalling observer will measure physical high energy quanta emerging from the black hole, and get burnt.
	\vspace{-3mm}
	\item If the observer gets burnt, then  we cannot have any complementary description where he falls through without noticing anything at the horizon.
\end{enumerate}

 From points (a) and (b) above, we find that the AMPS gedanken experiment  does not lead to the conclusions they suggest.   If one wishes to avoid Planck-scale physics, then one should restrict to measurements made  outside the stretched horizon. For such measurements
 point (b) shows that an infalling observer  will see the traditional black hole and a radiating stretched horizon as statistically similar systems.
  The underlying reason for this equivalence is that there is too little time for him to detect the Hawking quanta before he reaches the horizon. More importantly, point (a) shows that even if the infalling observer were to hit the stretched horizon violently, this fact would not by itself invalidate the possibility of complementarity; in  fact it is this very interaction that is expected to admit a complementary description.
  
 In the Discussion (Section \ref{secsix}) we summarize the essential physics involved in the conjecture of fuzzball complementarity to show precisely why it is not addressed by the AMPS argument. 
 
 The reader who is already familiar with fuzzballs and the conjecture of fuzzball complementarity may skip directly to Section \ref{secfour}.

\section{The information paradox and the fuzzball proposal}\label{basics}
 
 In this section we review the resolution of the information paradox through the fuzzball construction in string theory. Though the later arguments will be more abstract, the steps below will help us decide the validity of these arguments.

 \b
 
\noindent{ {\bf (a) The traditional black hole}}

 The information paradox arises from the way Hawking radiation is emitted from the {\it traditional black hole}. We define the traditional black hole as follows. There is a horizon, and a neighbourhood of the horizon with the following property. One can choose good slices in this neighbourhood, and in these good coordinates physics is `normal'. Here `normal' physics means exactly what we mean by normal physics in the lab: evolution of long wavelength modes ($\lambda\gg l_p$) is given by local quantum field theory on curved space, with corrections controlled by a small parameter $\epsilon$. These corrections can come from any quantum gravity effect, local or nonlocal, and all we require is that $\epsilon\r 0$ as $M\r \infty$, where $M$ is the mass of the black hole. 

\b

\noindent{ {\bf (b) The information paradox}}

 The traditional black hole arose from a study of gravitational collapse that leads to the Schwarzschild metric
\be
ds^2=-(1-{2M\over r})dt^{2}+{dr^2\over 1-{2M\over r}}+r^{2}{d\Omega_2^{2}}
\label{one}
\ee
If we use semiclassical gravity to follow the evolution of quantum modes during the collapse, we get the traditional black hole. We have  the a vacuum region around the horizon which indeed gives `lab' physics in a good slicing (i.e., in Kruskal coordinates). Evolution of vacuum modes at this horizon leads to entangled pairs being created, with  one member of the pair staying in the black hole and the other escaping to infinity as Hawking radiation. The entangled pair can be modeled for simplicity by \cite{cern}\footnote{Further analysis of such `bit models' can be found in \cite{plumpre,plumberg,bits,giddings}.}
\be
|\psi\rangle_{pair}={1\over \sqrt{2}}\left ( |0\rangle_{in}|0\rangle_{out}+|1\rangle_{in}|1\rangle_{out}\right )
\label{two}
\ee
The entanglement between the inside and outside grows by $\ln 2$ with each emission. Near the endpoint of evaporation this would leave just two possibilities: information loss or a remnant \cite{hawking,cern}. Both of these look unsatisfactory; we would like a pure state of Hawking radiation carrying all the information of the black hole. 

\b

\noindent{ {\bf (c) The theorem controlling small corrections}}

 The problem would be resolved if gravitational collapse led to a state other than the traditional black hole. But the traditional black hole solution appeared to admit no deformations, leading to the phrase `black holes have no hair'. Exactly the same problem holds for black holes in AdS. Thus AdS/CFT duality cannot by itself help to resolve the problem (for a detailed discussion of this issue, see \cite{cern,conflicts}).

This situation led many string theorists to the following belief. Hawking computed the pair creation at leading order, but there can always be small quantum gravity corrections to the wavefunction (\ref{two})
\be
|\psi\rangle_{pair}={1\over \sqrt{2}}\left ( |0\rangle_{in}|0\rangle_{out}+|1\rangle_{in}|1\rangle_{out}\right )+\epsilon {1\over \sqrt{2}}\left ( |0\rangle_{in}|0\rangle_{out}-|1\rangle_{in}|1\rangle_{out}\right )
\label{twenty}
\ee
where we have added a small amount of an orthogonal state for the pair. The correction $\epsilon$ for each pair must be small since the horizon geometry is smooth, but the number of emitted quanta is large ($\sim (M/m_p)^2$), and the net effect of the small corrections may accumulate in such a way that the overall state of the radiation would not be entangled with the black hole.

But in \cite{cern} it was shown that this hope is false; the change in entanglement $\delta S_{ent}$, compared to the entanglement $S_{ent}$ of the leading-order Hawking process, is bounded by
\be
{\delta S_{ent}\over S_{ent}}<2\epsilon \,.
\label{three}
\ee
This inequality is the essential reason why the Hawking argument has proved so robust over the years -- no small corrections can save the situation. We will make use of (\ref{three}) many times; many arguments in the other papers we discuss are also based on this inequality. 

\b

\noindent{ {\bf (d) The fuzzball structure of microstates}}

 In \cite{emission} it was found that a bound state in string theory {\it grows} in size with the number of branes in the bound state and with the coupling, so that its wavefunctional is always spread over a radius which is order the Schwarzschild radius. This growth in size is a very stringy effect; it arises from the phenomenon of `fractionation' \cite{dasmathur} which uses the extended nature of fundamental objects in the theory. Such horizon sized wavefunctionals are termed `fuzzballs'.

\begin{figure}[t]
\begin{center}
\includegraphics[scale=.4]{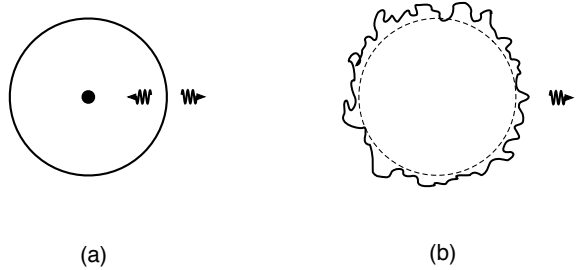}
\caption{(a) The traditional black hole; small corrections at the horizon {\it cannot} get information out in the Hawking radiation. (b) The fuzzball picture of black hole microstates; spacetime ends in stringy theory sources just before the horizon is reached. }
\label{fdiss}
\end{center}
\vspace{-2mm}
\end{figure}

The size of fuzzball states is estimated by using the entropy of brane bound states, together with the physics of fractionation. Thus this size estimate involves {\it all} the states of the black hole. To study the properties of fuzzballs further, it is useful to look at states where we place `many quanta in the same mode'. This is analogous to black body radiation, where placing a large number $N$ of quanta in the same harmonic gives a laser beam, with quantum fluctuations suppressed as $\sim {1\over N}$.\footnote{This study of low fluctuation states has led some to be confused about the nature of fuzzballs. They ask: are fuzzballs just solutions to supergravity or do they involve stringy degrees of freedom? As can be seen from the above discussion, there is no fundamental classical/quantum divide between states; all we can do is look at states with small or large fluctuations. In particular the non-BPS states studied in  \cite{ppwave} using the pp-wave technique were given in terms of strings placed in a fuzzball geometry. The correct question is not; `how messy is the fuzzball'; the only relevant question is `do we get a traditional black hole (with `lab physics' around a horizon) or do we not'. The only feature common to all fuzzballs is that we never form a traditional horizon.} One find that the fuzzballs generate a spacetime that resembles the traditional black hole far away from the horizon, but which ends\footnote{The word `end' should be understood as follows. In all known examples, individual black hole microstates are described by solutions of string theory involving smooth geometry far from the black hole, no horizon, and thus no interior (where `interior' refers to the space-time inside the horizon of the corresponding classical black hole solution). For generic states, the structure at the scale of the would-be horizon may be expected to have Planck-scale degrees of freedom (see also Footnote 4). In general, since there is no interior, we say that space-time ends outside the would-be horizon.} in a set of  string theory sources before reaching the horizon \cite{lm4,fuzzballs,fuzzball3}. This is pictured in Fig.\;\ref{fdiss}.\footnote{For the two-charge BPS black hole, all states have been shown to be fuzzballs. For other black holes, some fraction of the states have been constructed, and in each case have been found to be fuzzballs.}

\b

\noindent{\bf (e) Resolution of the paradox}

Given the existence of fuzzballs, the information paradox is resolved as follows. Fuzzballs do not radiate by pair creation from an `information-free horizon'; instead the radiation emerges from the surface of the fuzzball and carries information just like any normal body. This radiation has been explicitly worked out for simple fuzzballs; the rate of radiation agrees exactly with the Hawking emission rate expected for those fuzzballs but the details of the fuzzball state are seen to be imprinted in the spectrum of emitted quanta \cite{radiation}. 

If we start with a collapsing shell, then its wavefunction spreads over the enormous phase space of fuzzball states \cite{tunnel}, and then these fuzzball states radiate like any other warm body. The time for this spread can be estimated to be much smaller than the Hawking evaporation time \cite{rate}
\be
 t_{fuzzball}\ll t_{hawking}
 \label{four}
 \ee
This solves the information paradox.

\section{Traditional complementarity vs Fuzzball complementarity}\label{seccomp}

In this section we will explain what we mean by having a `complementary description'. We start by giving a toy example: the case of AdS/CFT duality \cite{maldacena}. This toy model is new. We briefly recall the traditional notion of complementarity, and then turn to how complementarity is conjectured to arise in the fuzzball description of microstates. This `fuzzball complementarity' has things in common to the toy example of AdS/CFT duality, but also differs from it in a crucial way.

\subsection{Toy example of complementarity: AdS/CFT duality}\label{ads}

We start with  an example that illustrates what we mean by having a complementary description. In this example an infalling quantum will encounter some degrees of freedom and appear to `go splat'; i.e. get `destroyed'. Yet there will be an alternative description where it continues unscathed. When a description of the latter kind exists, we will say that we have a `complementary' description of the degrees of freedom in the former description.

\begin{figure}[!]
\includegraphics[scale=.85]{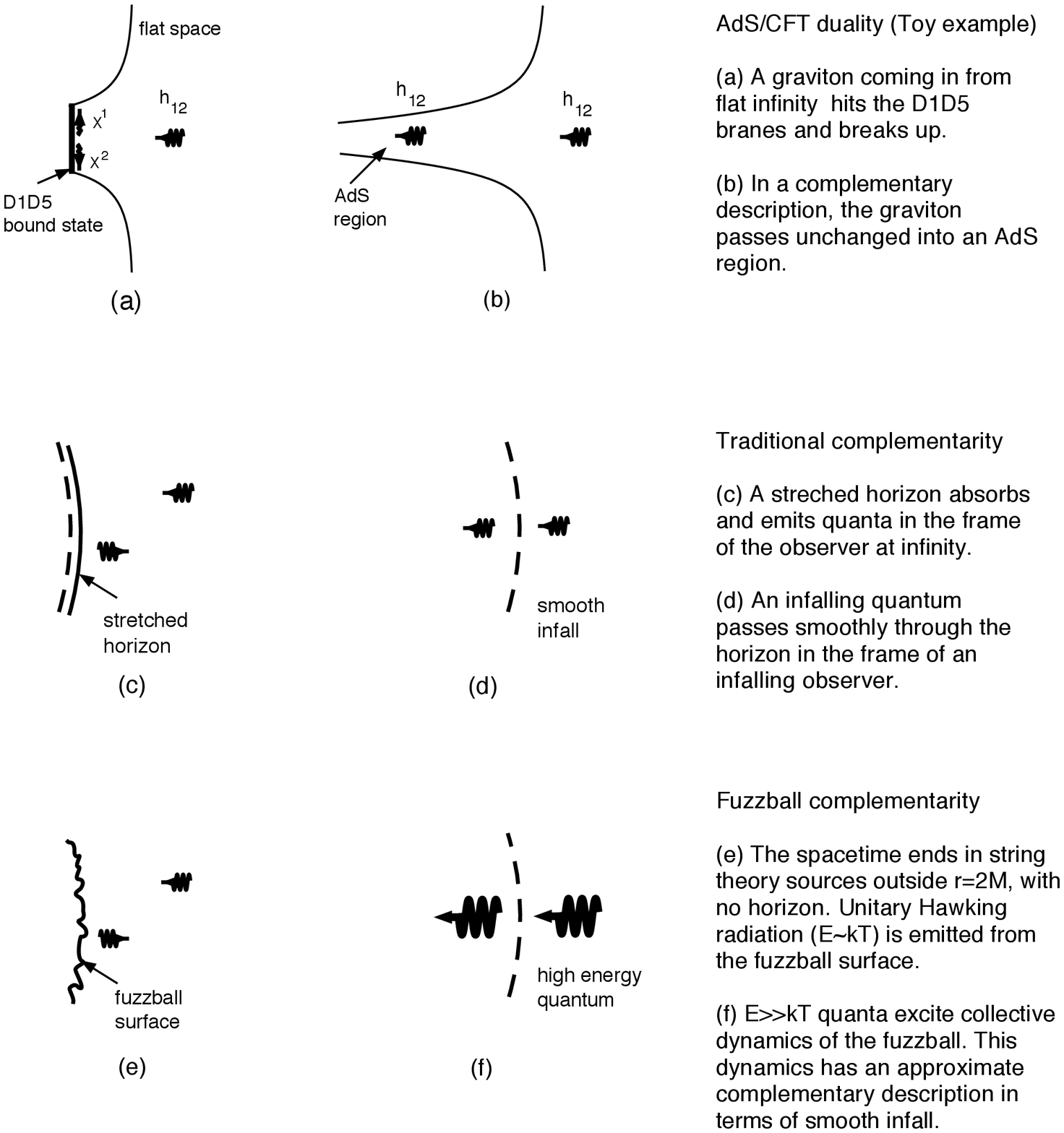}
\caption{AdS/CFT duality, traditional complementarity and fuzzball complementarity.}
\label{fz2p}      
\end{figure}

Consider IIB string theory compactified on $S^1\times T^4$. Let  $y$ be the coordinate along $S^1$ and $z_1, \dots z_4$ be the coordinates on $T^4$. 
We consider a bound state of $n_1$ D1 branes  wrapped on $S^1$ and $n_5$ D5 branes wrapped on $S^1\times T^4$. This bound state is depicted in Fig.\;\ref{fz2p}(a), where the direction along the branes is the $S^1$.

We are working in the context of a D-brane bound state in flat space, where in one description we have a CFT coupled to flat space, and in the other description we have a geometry with flat asymptotics and an AdS throat. The degrees of freedom deep inside the AdS throat (on the gravity side) will not play a role in the following.

To be more specific, we take the AdS radius $R_{AdS}$  to be macroscopically large. On the gravity side, we consider a throat which is very long in units of $R_{AdS}$  (measured by proper distance along a radial geodesic). We fix a CFT location $r=R_{CFT}$  in the usual way. We then consider the trajectory of an infalling quantum along a radial geodesic from say one AdS radius of proper distance above $r=R_{CFT}$  to one AdS radius of proper distance below this location (on the gravity side).

In the description involving a CFT coupled to flat space, the transition from an infalling graviton in flat space to CFT degrees of freedom is described by the corresponding CFT operator which describes the absorption (see e.g.~\cite{Avery:2009tu}).

A graviton with both indices on the $T^4$ is a scalar in the remaining dimensions. Consider in particular the graviton $h_{12}$, arriving at the brane bound state as shown in Fig.\;\ref{fz2p}(a).

In the CFT description, on hitting the brane bound state, the energy of the graviton gets converted to vibrations of the branes (open strings);\footnote{The actual evolution on the branes is more complicated when we consider interactions in the CFT, but  this simple picture illustrates the point we wish to make. Note that we are considering the gravity description at weak coupling, and so the CFT description is at strong coupling. But the important fact is that there are {\it two} descriptions at the same coupling; one using strongly interacting CFT of freedom, and one using the spin 2 graviton and higher closed string modes. In the former description the incoming $h_{12}$ appears to break up into pieces, while in the latter it remains intact.} a vibration polarized in the direction $X^1$ moves up along the $S^1$ and a vibration polarized in the direction $X^2$ moves down the $S^1$ \cite{comparing,interactions,malstrom}.

One may say that the graviton has `gone splat' on hitting the branes, to such an extent that it has split into two parts, $X^1$ and $X^2$. These two products obtained after impact certainly do not look like the single graviton $h_{12}$ that was arriving towards the brane bound state\footnote{At strong coupling, the graviton is absorbed into degrees of freedom of the strongly coupled CFT, where we cannot make precise statements. Nevertheless, the graviton may still be described as having `gone splat' in the sense that, in the CFT description, it is no longer a graviton and has been converted into strongly coupled CFT degrees of freedom.}. But as we well know, there is an alternative description of this physics where we replace the brane bound state by an AdS region.  In the latter description, the graviton $h_{12}$ falls smoothly into the AdS region, remaining as a single entity $h_{12}$ (Fig.\;\ref{fz2p}(b)). We can call this latter description a `complementary' description of the interaction with the D1D5 branes. Now we can ask our question: when the incoming graviton broke up on the D1D5 brane bound state, did it go `splat' or not?

To better understand how to interpret this situation, we look at a more detailed example where we start with {\it two} gravitons, $h_{12}$ and $h_{34}$, separated by a distance $D$. We can think of this pair of gravitons as being an `object'; if the separation of the gravitons is increased or decreased, we can say that the object has `been damaged' and `feels pain'. 

At zero coupling in the CFT, the evolution of these gravitons proceeds as follows \cite{lm4}. First $h_{12}$ hits the D1D5 bound state, and changes to excitations $X^1, X^2$ which travel at the speed of light  in opposite directions along $y$. At a later time $h_{34}$ hits the bound state and changes to vibrations $X^3, X^4$, again separating at the speed of light. But the separation $D$ between the initial gravitons can be recovered from the open string excitations. Let $y_i$ be the location along the $S^1$ of the excitation $X^i$, for $i=1,2,3,4$. We have
\be
y_1=t, ~~y_2=-t, ~~y_3=t-D, ~~y_4=-t+D
\ee
so the value of $D$ is encoded in the vibrations $X^i$ as
\be
D=\h[(y_1-y_2)-(y_3-y_4)]
\ee
In the dual gravity description, the two gravitons fall  smoothly into the AdS, maintaining their separation $D$ and thus showing no indication of `damage' or `pain'.  But given that the CFT description is a faithful copy of the gravity description, and that we can recover the same value $D$ from the CFT, it looks correct to say that there is no damage or pain felt  in the brane description either.\footnote{Again, to make the toy example more accurate, one should consider the CFT at strong coupling. The basic result is unchanged: in the CFT description the incoming graviton is absorbed into degrees of freedom of the strongly coupled CFT, which in no way resemble the incoming gravitons, yet which somehow encode the value of $D$.}

By contrast, when we throw an object onto  a normal concrete wall, we do not expect to find a complementary description. Let us analyze what was special about the D1D5 brane case which did allow for complementarity.

In the D1D5 example, the Hilbert space of the incoming gravitons mapped faithfully into the Hilbert space of vibrations of the branes. That is, if we write the eigenstates of the incoming graviton as $|\psi_E\rangle$ and the eigenstates of the D1D5 system as $|\t\psi_E\rangle$, then we find
\be
\int\,  dE \, C(E)\,  |\psi_E\rangle ~\r~ \int \, dE \, C(E)\,  |\t \psi_E\rangle
\label{eone}
\ee
 The {\it nature} of the excitations changed completely - they changed from being gravitons to being vibrations of branes - but this is {\it not} important. What is important is that the amplitude for a given energy remained the same (or approximately the same). A important input for getting a relation like (\ref{eone}) is that   the D1D5 bound state had a very closely spaced set of energy levels.  This high density of levels leads to a `fermi-golden-rule' absorption of the graviton, and in such an absorption each incoming energy level $E_k$ transfers its amplitude to energy levels $\t E_k$ that are very close to $E_k$. (In \cite{interactions} the absorption of the graviton onto the brane bound state was computed  by such a fermi-golden rule process.)

What {\it does} cause `damage' or `pain' is the situation where the levels available in the absorbing system are not sufficiently continuous. In this situation we will find in general 
\be
\int \,  dE \, C(E)\,  |\psi_E\rangle ~\r~ \int \, dE \, C'(E) \, |\t \psi_E\rangle, ~~~C(E)\ne C'(E)
\label{eoneq}
\ee
In particular, a concrete wall will not have  the same energy levels as the object hitting it, and so the incoming object will not be mapped faithfully into excitations of the concrete wall. In this situation we do not expect a complementary description of the impact. 

To summarize, we cannot just say: `If we go `splat' on hitting some degrees of freedom, then we cannot have complementarity'. The impact transfers excitation energy to the degrees of freedom that are encountered. To know if we can have a complementary description we have to ask if the Hilbert space of the infalling object  maps faithfully into a subspace of the Hilbert space of the encountered degrees of freedom. 

\subsection{Traditional  complementarity}\label{trad}

In the early works on black hole complementarity \cite{complementarity}, the physics that was proposed is depicted in Fig.\;\ref{fz2p}(c),(d). It was assumed that we can place a `stretched horizon' just outside $r=2M$, and that incoming quanta could be taken to interact with degrees of freedom on this stretched horizon. In the complementary description, we have just the smooth infall through the horizon.

The problem with this proposal is discovered when we ask for the physical origin of the degrees of freedom on the stretched horizon. It was argued that since the Schwarzschild coordinates break down at $r=2M$, there will be violent fluctuations of the gravitational degrees of freedom as we approach $r=2M$. It was further argued that these violent fluctuations are indicative of the fact that physics outside the horizon is self-consistent, and the stretched horizon provides the natural boundary beyond which we need not look.

Such an argument is, however, unsatisfactory. The breakdown of Schwarzschild coordinates means that we should use better coordinates, not that we are entitled to assume new physics. But there is an even more serious difficulty with this proposal, which we can see by returning to our basic question: how does the information paradox get resolved? There is a `smooth slicing' of the geometry where we see the creation of entangled pairs (\ref{two}). The defenders of  traditional complementarity  argued that the inner and outer parts of the horizon should not be considered in the same Hilbert space, since an observer who falls in has strong limitations on how he can communicate with the outside; thus the state (\ref{two}) makes no sense.  But no mechanism was proposed to implement such a drastic change to normal physics. The skeptics of complementarity simply noted that there {\it is} a good slicing of the geometry which we should use to do physics at the horizon, and with this slicing there appears to be no reason to not have a single Hilbert space that includes both the inner and outer parts of the horizon. 

For these reasons, the traditional picture of complementarity remained an unresolved issue.  It is important to note the difference between the traditional black hole case and the example of AdS/CFT that we presented in Section \ref{ads}. In the AdS/CFT example of Fig.\;\ref{fz2p}(a),(b), the boundary where we get a complementary description is not a horizon, and there is no particle creation there. Thus we do not have the information problem. But in the case of a black hole there is no way to stop the creation of entangled pairs in any picture where a smooth horizon is assumed, and then we cannot scape the information paradox. As we will see now, the way complementarity can arise with the fuzzball picture in string theory is somewhat different, and needs us to recognize that real degrees of freedom appear at the location of the horizon. 

\subsection{The proposal of fuzzball complementarity}

With the explicit construction of black hole microstates in string theory (fuzzballs) we find that things work out  differently from the traditional picture of complementarity. The general idea of `fuzzball complementarity' is developed in \cite{plumpre,plumberg,otherfcrefs}. The notion of making spacetime by entanglement \cite{raamsdonk,israel,eternal} is very useful in this approach.   Here we just give an outline of how things work:

\parskip=10pt

(a) Complementarity does {\it not} arise because of a choice of coordinates (Schwarzschild vs Kruskal). Instead, the construction of microstates is fully covariant.

(b) In the traditional black hole we have {\it vacuum} around the horizon. But in string theory, spacetime has a `boundary' where it ends with in a set of string theory sources just outside $r=2M$, before the horizon is reached. The details of these sources encode the choice of microstate. 

(c) Hawking radiation arises as quanta radiated from the details of microstate structure near the boundary. For simple microstates this radiation has been explicitly computed, and it arises from `ergoregion emission' \cite{radiation} near the boundary. The details of the ergoregion structure depend on the choice of microstate.

(d) Since we have `real' degrees of freedom at the horizon, the $E\sim kT$ quanta radiated from the microstate are able to carry out the information of the microstate. We {\it cannot} have a complementary picture where we replace the physics of such quanta by the vacuum physics seen at the horizon of the traditional black hole. In this way our complementarity differs from traditional complementarity.
What we have to do is make a distinction between $E\sim kT$ quanta (relevant for the information problem) and $E\gg kT$ quanta (relevant for the `infall problem' of heavy observers). It was conjectured in~\cite{plumberg} that the complementary description should describe measurements in the frame of a lab (composed of $E\gg kT$ quanta) falling freely from infinity to the surface of the fuzzball. We can describe such a process as a `hard-impact' process.

(e) Let us restate the previous point another way. In the fuzzball scenario, the exact state near the horizon is not the vacuum state of an infalling observer, or anything close to it; it is expected to have Planck-scale degrees of freedom. Thus we cannot say that we have low energy effective field theory at the horizon, and then use this low energy field theory for the purpose of describing all possible low energy observations of an infalling observer. Instead, we conjecture a complementary description for hard-impact processes involving $E\gg kT$ quanta.

(f) The complementarity conjecture is now the following (Fig.\;\ref{fz2p}(e),(f)). 
Given a hard-impact process involving $E\gg kT$ quanta, the resulting dynamics can be reproduced to a first approximation by the geometry of the black hole interior, for times of order crossing time (i.e. before the quanta reach the singularity). This description emerges from the fuzzball dynamics  as follows. The $E\gg kT$ quanta excite collective modes of the fuzzball. To a  first approximation, the evolution of these modes is insensitive to the precise choice of fuzzball microstate (assuming we have taken a generic microstate). The evolution of these collective modes in this leading approximation is to be encoded in the complementary description. Thus, let the initial state of the hole have mass $M$ and be the linear combination of fuzzball states $\sum_i C_i |F_i\rangle$. When a quantum of energy $E\gg kT$ impacts hard onto the fuzzball surface, the wavefunction of the fuzzball shifts to a combination $\sum_j C'_j F'_j$ over the fuzzball states with mass $M+E$:
\be
\sum_i C_i |F_i\rangle~\r ~\sum_j C'_j F'_j
\label{evolve}
\ee
If $E\gg kT$, then the number of coefficients $C'_j$ is much larger than the number of $C_i$. The leading order evolution of the coefficients $C'_j$ is to be captured by the complementary description. 

(g) We can now see the similarities and differences with the toy example of AdS/CFT duality discussed in Section \ref{ads}: \vspace{-5mm}
\begin{enumerate}[(i)]
	\item The D1D5 brane degrees of freedom are analogous to degrees of freedom at the `boundary' of the fuzzball microstate. \vspace{-2mm}
 \item The D1D5 branes were taken to be in their  ground state,\footnote{We can take excited states of the D1D5 branes, but in AdS/CFT duality we take these to be low energy excitations, and their effect in the dual gravitational description will occur near $r=0$, not near the place where the CFT is placed.} while the fuzzball structure differs microscopically from state to state. Thus we get only approximate complementarity in the black hole case, by looking at hard-impact, $E\gg kT$ processes where the details of the fuzzball microstate become irrelevant. \vspace{-2mm}
 \item In the AdS/CFT case the complementary description was possible because of the closely spaced levels of the D1D5 brane system. In the black hole case we again have a close spacing of levels, which is guaranteed by the large number $Exp[S_{bek}]$ of fuzzball microstates. 
\end{enumerate}

\parskip=3pt

\subsection{Summary}

To summarize, we have observed the following:

(a) In our toy example of AdS/CFT duality, we have a brane description, where an incoming quantum appears to hit some degrees of freedom violently and `break up'. In a `complementary description', the incoming quantum smoothly through into an AdS region. There is no radiation from the AdS boundary itself, so there is no creation of entangled pairs at that location.

(b) In traditional complementarity, one argues that there are two equivalent descriptions, a fact allowed by the limitations on communication between observers inside and outside the hole.  In one description (that of the outside observer)
incoming quanta are reflected back as Hawking radiation from a stretched horizon, while in another description (that for an infalling observer) the horizon is a smooth place. Since there is a horizon, there is a creation of entangled pairs (\ref{two}) in a smooth slicing at that location, and there is no clear mechanism to remove this entanglement.

(c) In fuzzball complementarity, there are real degrees of freedom at the horizon which arise from the fact for each black hole microstate, the compact directions pinch off in a mess of string sources and spacetime ends before we reach $r=2M$. The details of this `fuzzball' differs from microstate to microstate; there is no Hawking type creation of entangled pairs and the radiation from the fuzzball surface can be explicitly seen carry information of the microstate. Since the fuzzball surface differs from microstate to microstate, complementarity can only be obtained in an approximation where the effect of these  differences is small. The conjecture is that when  $E\gg kT$ quanta impact the fuzzball, they  excite collective modes that are relatively insensitive to the precise choice of microstate;  the  evolution of these modes (\ref{evolve})
 can be approximated by evolution in a spacetime that mimics the black hole interior. 
 
\section{Limits on measurements made outside the horizon}\label{secfour}

In this section we address the following question. If we measure the radiation outside a black hole, then can we tell the difference between a traditional black hole and an object that radiates unitarily at the same temperature $T$ from a surface just outside $2M$?

\begin{figure}[htbt]
\begin{center}
\includegraphics[scale=.75]{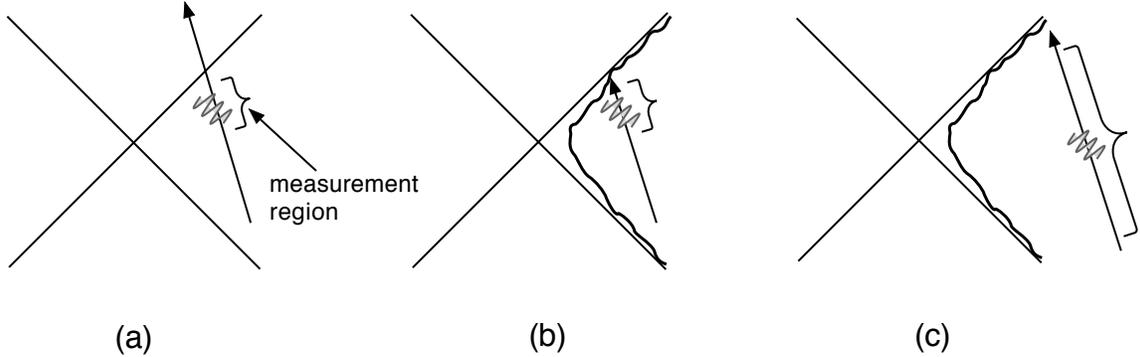}
\end{center}
\caption{(a) An inertial detector in Minkowski space, making a measurement using only the indicated part of its trajectory. Vacuum fluctuations excite the detector. (b) A similar detection, but for case of a warm body radiating into the right Rindler wedge. The wavelength of quanta is of the same order as the distance from the horizon. (c) Radiation from a `hot' body, where the wavelength is much shorter than the distance from the horizon.}
\label{fz6}     
\end{figure}

The measurements we are interested in are close to the horizon ($|r-2M|\ll 2M$), so we can consider the near horizon geometry depicted in Fig.\;\ref{fz6}. In Fig.\;\ref{fz6}(a) we have the traditional black hole, which has vacuum around the horizon, so the near horizon region looks like  Minkowski space when seen in Kruskal coordinates. In Fig.\;\ref{fz6}(b) we have a warm surface placed just outside $r=2M$ (indicated by the jagged line), and this surface is assumed to radiate quanta at the temperature $T$ of the black hole.

At first it may appear that the case of Fig.\;\ref{fz6}(b) has real radiation that can `burn', while there is no real radiation in Fig.\;\ref{fz6}(a). We {\it can} see quanta in Minkowski spacetime by taking a detector that accelerates. But our interest in in freely falling observers, which are indicated by the straight line trajectory in Fig.\;\ref{fz6}(a). One may expect  that a detector moving in straight line in Minkowski space should not detect any quanta. But the situation we have is a little special. We are asking if we can distinguish the physical situations of Fig.\;\ref{fz6}(a) and Fig.\;\ref{fz6}(b) by observations {\it outside the horizon}. Thus a detector trying to make a measurement would have to do this task by using only a section of its trajectory like that indicated in Fig.\;\ref{fz6}(a).

But if we place conditions on how long a detector has to make a measurement, then we run into the problem that we pick up vacuum fluctuations. We discuss the scales involved in the problem in Section \ref{sectime}. Suppose we are considering radiation at the Hawking temperature $T$.  The wavelength of these quanta at a distance $d$ from the horizon is $d\sim \lambda$. The infalling detector trying to measure such quanta has a limited time to make this measurement, and we argue that this available time is less than the time required to make the desired measurement.

In Section \ref{rindler} we note that the above estimates reflect a general fact: for generic states of the radiating body in Fig.\;\ref{fz6}(b), observations of radiation do not appear statistically different from the vacuum fluctuations picked up by the detector of Fig.\;\ref{fz6}(a). The arguments we give are very basic to the theory of particle detection, and are implicit in many treatments of Rindler space (see e.g. the review \cite{Crispino:2007eb} and references within).\footnote{We also thank Bill Unruh for an earlier conversation about detectors in Minkowski spacetime.}

\subsection{Time needed for detector response}\label{sectime}

Let us  examine what kind of quanta a detector can actually pick up in a measurement process. In Appendix \ref{detect} we show that if we wish to measure a quantum of wavelength $\lambda$, the we need a proper time $\gtrsim\lambda$ to elapse along the detector trajectory:
\be
\Delta \tau_{needed}\gtrsim \lambda
\label{ten}
\ee
 In Fig.\;\ref{fz6} we note two different possibilities for the location of the quantum of wavelength $\lambda$. In Fig.\;\ref{fz6}(a),(b) the quantum is at a distance $d\sim \lambda$ from the black hole surface. In Fig.\;\ref{fz6}(c) the quantum is at a distance $d\gg \lambda$ from the black hole surface. In Appendix \ref{wavelength} we show that the Hawking quanta radiated from the black hole surface are of the former type; the typical wavelength found at a distance $\lambda$ from the horizon is $\sim\lambda$ itself:
 \be
 \lambda \sim d
 \label{el}
 \ee
  We now begin the see the source of difficulty in catching high energy Hawking quanta: we are already very close to  the horizon when we encounter them, and then we may have too little time left to interact with them. Before proceeding, there is one effect that we must take into account. Because the detector is infalling, it sees the outgoing quantum as being Lorentz contracted; thus the wavelength of the quantum appears shorter than the distance $d$ measured along a $t=const$ slice. We take a local Lorentz frame oriented along the Schwarzschild $t, r$ directions, and let the proper velocity of the detector in this frame be
  \be
U^{\hat t}=\cosh\alpha, ~~~U^{\hat r}=-\sinh\alpha
\ee
Then, as shown in Appendix \ref{wavelength}, the effective wavelength of the Hawking quanta encountered by the infalling detector is
\be
\lambda_{eff}\sim d e^{-\alpha}
\ee
Now we consider the proper time available to an infalling detector to measure the Hawking quantum; this detection must be made between the time the detector is at a distance $\sim d$ from the horizon and the time it falls through the horizon. In Appendix \ref{time} we show that for a detector falling in from far outside the horizon, this proper time is
  \be
  \Delta \tau_{available}< de^{-\alpha}
  \label{tw}
  \ee
  Putting together (\ref{ten}), (\ref{el}) and (\ref{tw}) we get
  \be
   \Delta \tau_{available}< \Delta \tau_{needed}
   \ee
   so we conclude that an infalling detector cannot reliably pick up Hawking quanta being radiated from a black hole surface. We now turn to comparing the behavior of detectors in the situations of Fig.\;\ref{fz6}(a) and Fig.\;\ref{fz6}(b).

\subsection{Detectors in Rindler space and detectors near warm bodies}  \label{rindler}

Let us consider the following question. We look at the situation of Fig.\;\ref{fz6}(a), where we have an inertial detector in empty Minkowski space, but the detection is required to be made before the detector crosses the Rindler horizon. We can therefore capture our physics by using  Rindler coordinates covering the right Rindler wedge 
\be
t_M=r_R \sinh t_R, ~~~x_M=r_R \cosh t_R
\ee
where $t_M, x_M$ are the Minkowski coordinates and $r_R, t_R$ are the Rindler coordinates. Now consider the behavior of the detector as seen in these Rindler coordinates. The space near the horizon looks very hot; it is full of Rindler quanta. Would these quanta `burn' the infalling detector?

At first one may think that an inertial detector in Minkowski space should see nothing. But we have already noted above that the limits placed on the measuring time causes the detector to be excited by vacuum fluctuations. We will now see that such an excitation is of the same kind as that expected in Fig.\;\ref{fz6}(b), where we have `real' quanta being radiated at the Rindler temperature by a surface placed just outside the Rindler horizon. 

Let the quanta being detected correspond to a scalar field $\phi$, which is taken to be in the Minkowski vacuum state $|0\rangle_M$. Since our observations are confined to the right Rindler wedge, we can use the expansion of the field operator in Rindler modes
\be
\hat \phi=\sum_\omega [f_\omega(r_R)e^{-i\omega t_R}\, \hat a_\omega+
f^*_\omega(r_R)e^{i\omega t_R}\, \hat a_\omega^\dagger]
\ee
Let the detector be a 2-level system. We will take it to start in the  unexcited state $|i\rangle$, and interactions with $\phi$ can move it to the state  $|f\rangle$. The interaction is described by $\int d\tau \, \hat H_{int}(\tau)$ where (see e.g.~\cite{Crispino:2007eb})
\be
\hat H_{int}(\tau)=q \, h(\tau)\, \hat O(\tau)\,  \hat \phi  \bigl(  t_R(\tau), r_R(\tau)\bigr)
\label{qint} 
\ee
Here $\hat O$ is an operator made out of the detector variables, $q$ is a coupling constant and $0\le h(\tau) \le 1$ is a `switching function' that allows us to switch on and switch off the interaction of the detector with the scalar field $\phi$. 

The Minkowski vacuum $|0\rangle_M$ can be written in terms of Rindler states of the left (L) and right (R) wedges
\be
|0\rangle_M=C\sum_k e^{-{E_k\over 2}}|E_k\rangle_L|E_k\rangle_R, ~~~~~~~C=\Big (\sum_k e^{-E_k}\Big )^{-\h}
\label{split}
\ee
Now suppose the interaction is switched on for a brief period as indicated in Fig.\;\ref{fz6}(a).  Before the interaction is switched on, the state of the overall system is
\be
|\Psi\rangle_i=|i\rangle \otimes C\sum_k e^{-{E_k\over 2}}|E_k\rangle_L|E_k\rangle_R
\label{qstate}
\ee
Using first order perturbation theory in the strength of the interaction $q$, we ask for the amplitude for the transition
\be
|i\rangle\otimes |E_k\rangle_L|E_k\rangle_R~\r~|f\rangle\otimes |E_k\rangle_L|E_{k'}\rangle_R
\ee
This amplitude is
\be
{\cal A}_{kk'}=-i \int_{-\infty}^\infty d\tau\,  h(\tau)\,  \langle f | \hat O |i\rangle {}_R\langle E_{k'} |\hat\phi \bigl( t_R(\tau), r_R(\tau)\bigr )|E_k\rangle_R
\ee
The quantity ${}_R\langle E_{k'} |\hat\phi \bigl(t_R(\tau), r_R(\tau)\bigr)|E_k\rangle_R$ can be easily computed by writing $|E_k\rangle_R$ in terms of the occupation numbers for different Rindler modes and using the field expansion (\ref{split}). Note that $h(\tau)$ in nonzero only over the part of the detector trajectory indicated in Fig.\;\ref{fz6}(a). 

The probability for the detector to get excited $|i\rangle\r |f\rangle$ is then
\be
P_{Minkowski}=|C|^2\sum_k e^{-E_k}\sum_{k'}|{\cal A}_{kk'}|^2
\ee
where the subscript on $P$ indicates that this computation was performed for the Minkowski vacuum situation of Fig.\;\ref{fz6}(a). Here the factor $e^{-E_k}$ reflects the fact that the probability of finding the state $|E_k\rangle_R$ in the state (\ref{split}) is
\be
p_{E_k}=|C|^2e^{-E_k}
\label{wone}
\ee

Now consider a state that describes a warm body at the same temperature as Rindler space, as shown in Fig.\;\ref{fz6}(b). In terms of Rindler eigenstates, this state has a form
\be
|\Psi\rangle=\sum_k C_k |E_k\rangle
\label{stateq}
\ee
Different microstates of the warm body have different coefficients $C_k$, but the ensemble average over possible microstates will have
\be
\langle |C_k|^2\rangle=|C|^2 e^{-E_k}
\label{approximation}
\ee
in agreement with (\ref{wone}).
We again consider the infalling detector with the same interaction (\ref{qint}). With the state (\ref{stateq}) the probability for the detector to get excited is
\be
P_{microstate}=\sum_k  \sum_{k'}|C_k|^2| {\cal A}_{kk'}|^2
\ee
Using (\ref{approximation}) we find that the the ensemble average of the excitation probability for radiation from `warm bodies' is the same as the excitation probability in the Minkowski vacuum when the detection range is confined to  be outside the horizon
\be
\langle P_{microstate}\rangle=P_{Minkowski}
\ee
In particular, if the infalling body is macroscopic so that it `measures' a large number of quanta, then the effect of radiation in any one microstate will be approximately the same as the effect of vacuum fluctuations when 
we restrict to the part of the observer worldline that is outside the horizon:
\be
 P_{microstate}\approx P_{Minkowski}
 \label{eqburn}
\ee
A similar effect is also obtained when we consider a detector that has fallen in from near infinity. Quanta at infinity with wavelength $\lambda$ are wavepackets that have a transverse size $\Delta \gtrsim\lambda$; this is necessary since otherwise the uncertainty principle will give the quantum more transverse momentum $\sim 1/\Delta$ than radial momentum, and the quantum will not really be headed towards the black hole. As the quantum comes closer to the horizon, the wavelength in the radial directions becomes small by blue-shifting, while the transverse size $\Delta$ remains unaffected. Thus all quanta falling in from infinity are `flattened' near the horizon. The largeness of $\Delta$ compared to the radial wavelengths of Hawking quanta near the horizon means that several Hawking quanta at different angular positions along the horizon can interact with the infalling quantum. Thus we are again led to compute statistical averages, getting a result like (\ref{eqburn}).

\subsection{Summary}

To summarize, we have compared measurements made by an infalling detector in the case of Minkowski space (Fig.\;\ref{fz6}(a)) and in the case of a warm body at the same temperature (Fig.\;\ref{fz6}(b)). These two cases are equivalent to the traditional black hole and to a black object with a radiating surface just outside the horizon. While one might at first think that the detector would measure very different things in the two cases, we find that the detector excitation probabilities are actually {\it similar}. The underlying reason for the similarity is the fact that we need the detection to be completed before the detector reaches the horizon, and this causes vacuum fluctuation excitations in the Minkowski space case that resemble the `real' quanta picked up in the warm body case.

While fuzzballs radiate at exactly the rate expected for Hawking emission, one may envisage a theory other than string theory where the quanta are emitted with energy 
\be
E\gg kT
\ee
with $T$ the Hawking temperature. In other words, we may give up the thermal spectrum of emission, and have the situation pictured in Fig.\;\ref{fz6}(c) where the emitted quantum has wavelength $\lambda\ll d$ at a distance $d$ from the horizon. In this case it {\it is} possible to make a reliable measurement of the quantum, since ample time is available before the detector reaches the black hole surface.  But in this case the emitted radiation will not carry away all the information of the black hole. This follows because the entropy of Hawking radiation (at temperature $T$) is just $\sim 1.3$ times the Bekenstein entropy $S_{bek}$ \cite{zurek}. Taking $E\gg kT$ will give us $N\ll S_{bek}$ quanta to carry out the $S_{bek}$ bits in the black hole, and this is not possible since each quantum carries $\sim 1$ bit of information.


\section{The AMPS argument}\label{secfive}

In this section we examine the main argument of Almheiri, Marolf, Polchinski and Sully (AMPS). We will note that the measurement they envisage cannot be performed reliably in the given situation, and further, that no conclusions about fuzzball complementarity can be drawn from such a situation.

In outline, the AMPS argument goes as follows: 
\begin{enumerate}[(i)]
	\item If Hawking evaporation is unitary, then the state near the horizon is not the vacuum in an infalling observer's frame, but involves high-energy excitations.
	\vspace{-3mm}
	\item If there are high-energy excitations near the horizon, then an  infalling observer will measure physical high energy quanta emerging from the black hole, and get burnt.
	\vspace{-3mm}
	\item If the observer gets burnt, then  we cannot have any complementary description where he falls through without noticing anything at the horizon.
\end{enumerate}

We examine each of these steps in turn.

\b

\noindent{ {\bf (i) The need for large corrections at the horizon}}

\begin{figure}[t]
\begin{center}
\includegraphics[scale=.75]{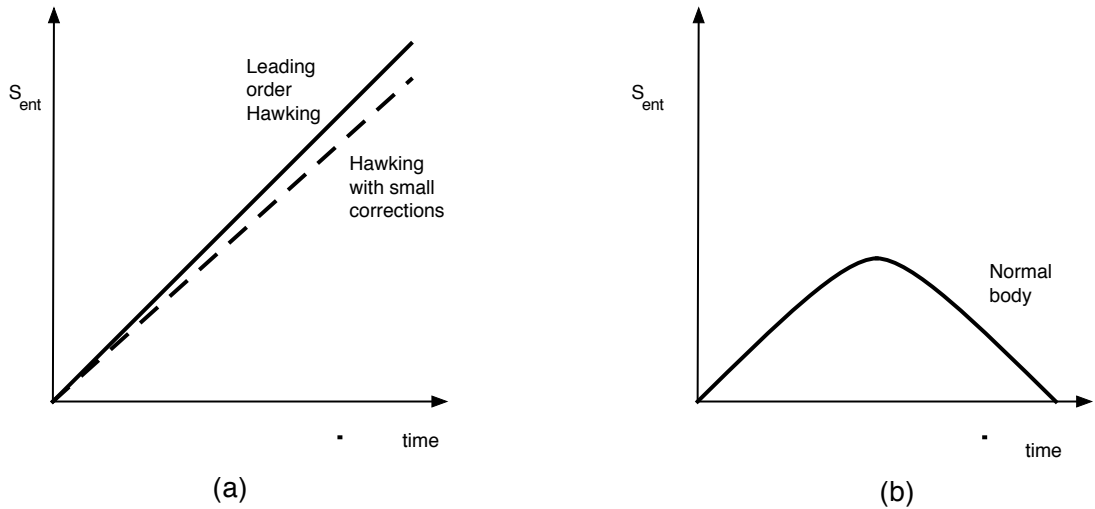}
\end{center}
%
%
\caption{(a) The growth of entanglement entropy for the traditional black hole in the leading order Hawking computation (solid line), and with small corrections allowed (dashed line). (b) The entanglement entropy expected for a normal body \cite{page}; $S_{ent}$ must return to zero when the body radiates away completely.}
\label{fz9}       
\end{figure}

 In \cite{cern} it was shown, using strong subadditivity, that semiclassical physics at the horizon cannot lead to the behavior of entanglement entropy $S_{ent}$ that is expected for normal bodies \cite{page}. The behavior for $S_{int}$ is depicted in Fig.\;\ref{fz9}. AMPS try to summarize a version of this argument, but miss a crucial step. We would like to clarify this point since it is important, before continuing with the AMPS argument.

\begin{figure}[htbt]
\begin{center}
\includegraphics[scale=.75]{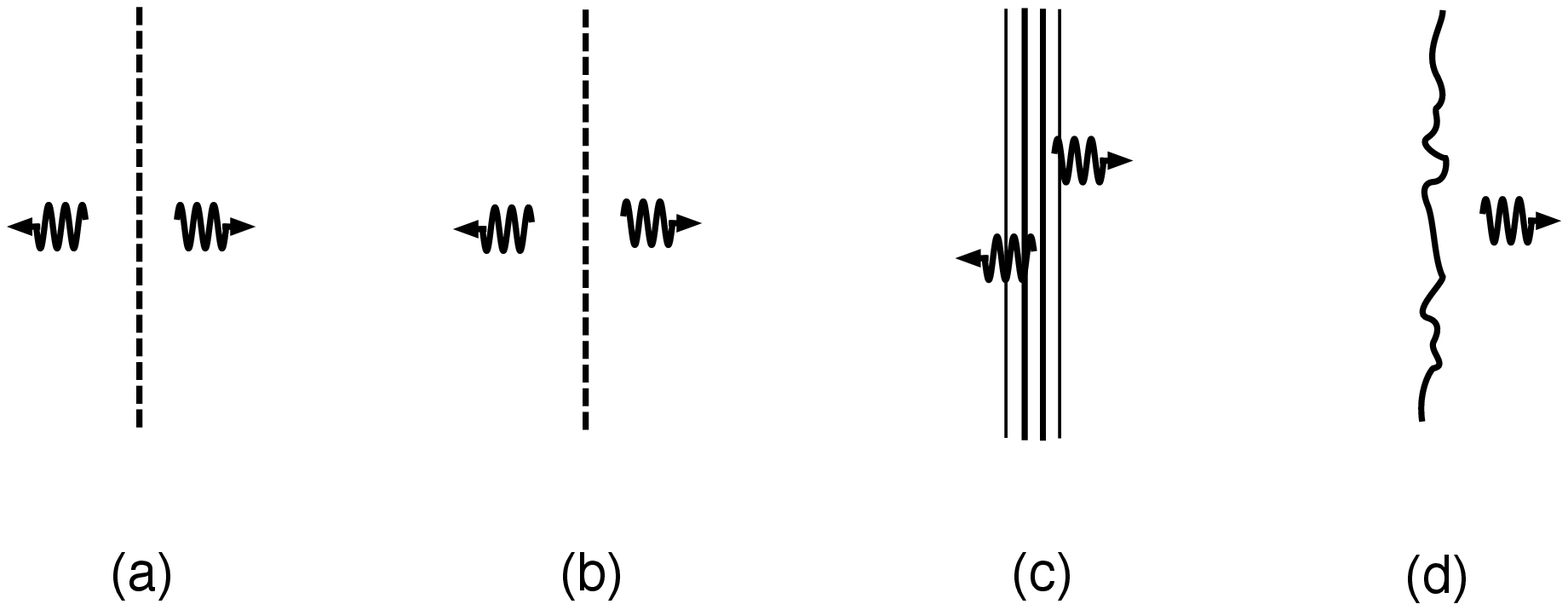}
\end{center}
%
%
\caption{(a) Creation of entangled pairs in the leading order Hawking computation. (b) Small corrections; if these could reproduce the graph Fig.\;\ref{fz9}(b) then we would not need a firewall. (c) A firewall that one can pass through; now one can detect the quanta near the horizon. (d) In a fuzzball spacetime ends before the horizon. Hawking radiation is an integral part of the dynamics of the fuzzball.}
\label{fz7}       
\end{figure}

Consider the Hawking pair (\ref{two}) produced in the leading order Hawking process; let the outer and inner members of this pair be called $B, C$ respectively. AMPS consider this leading order process, a fact which is implicit in their assumption that
$S_{BC}=0$;
i.e., the produced pair is not entangled with anything else (Fig.\;\ref{fz7}(a)). They then use strong subadditivity to argue that $S_{ent}$ cannot return to zero like it should for normal bodies. But this situation does not need the powerful relation of strong subadditivity. In the leading order Hawking process the relation (\ref{two}) tells us that the state of the created pairs is a tensor product of individual pairs (eq. (17) of \cite{cern}), and so $S_{ent}=N\ln 2$ after $N$ pairs have been produced. This gives the linearly increasing graph of Fig.\;\ref{fz9}(a), and we do not need strong subadditivity to prove that $S_{ent}$ does not return to zero.

The important issue, as discussed in Section \ref{basics}, is whether {\it subleading} corrections to the leading order Hawking process can make $S_{ent}$ reproduce the behavior of a normal body.\footnote{The possibility that this might happen was raised in \cite{eternal}. Hawking's reversal of his belief that information is lost was also implicitly based on the  assumption that exponentially small corrections to the leading order  process would produce an unentangled state \cite{hawkingreverse}.}  If small corrections could do the job, then we {\it cannot} conclude that   there would be a firewall; we depict this in Fig.\;\ref{fz7}(b). To analyze small corrections we have to start from 
$S_{BC}=\epsilon$
and then we do need to use strong subadditivity
 to establish  the required inequality (\ref{three}).

To summarize, a smooth vacuum at the horizon leads to the creation of Hawking pairs (\ref{two}), and with (\ref{three}) we see that we cannot get information out in Hawking radiation. Thus if we do wish to have the radiation be unitary, then we must alter the structure of the modes involved in the Hawking process.  One may try to restrict the required change to just these modes; this requires us to invoke as yet undiscovered  nonlocal effects \cite{giddings}. If we choose to not do this, then we have an alteration of the physics for {\it all modes} at the horizon. AMPS take the latter route\footnote{They consider the possibility of nonlocal effects in a separate discussion later.}, and then consider an experiment: they let an infalling observer fall into such a hole and argue he will get `burnt' by the altered structure at the horizon. Further, they argue that getting burnt in this way precludes the possibility of complementarity. We now examine each of these issues in turn.

\b

\noindent{ {\bf (ii) Getting burnt by Hawking quanta}}

Here AMPS wish to  distinguish the traditional black hole from a body that radiates at the Hawking temperature from a surface just outside the horizon. They argue that in the case of the radiating body an infalling observer will observe high energy quanta, while there will be no such quanta observed for the traditional black hole.  Let us see what  questions we can ask:

\b

(a) The temperature of the radiation is $T\sim m_{p}$ at a distance $l_{p}$ from 
the horizon. If we wish to avoid Planck-scale physics, the we can try to  focus on the radiation a distance $d$ from the horizon with
\be
l_p\ll d\ll 2M
\ee
For concreteness, let us think of $d\sim 10^6 l_p$, where we expect the temperature to be high enough to `burn' but the physics is still not the unknown physics at Planck scale. For instance, in Fig.\;1 of \cite{amps}, one can ask if the infalling detector gets burnt during the part of its trajectory where it crosses the shaded region representing the outgoing Hawking quantum.  To restrict our question to the required region, we consider the effect of the radiation on a detector which is switched off when we get to a distance closer than $\sim 10^6 l_p$  from the horizon. But in such a situation we have noted in Section \ref{rindler} that the excitation probability of the detector in the case of the traditional horizon and in the case of the firewall are statistically  the {\it same} (eq.(\ref{eqburn})). Thus we cannot say that we will get burnt in one case and not the other.

(b) The above equality of excitation probabilities resulted from the fact that we were allowed a limited time to make the detection; thus vacuum fluctuations excited the detector even for the traditional hole. We could allow ourselves a longer time for detection if we assumed that we could pass through the firewall to the other side of the horizon, and again find ourselves in a region of low temperature. Such a possibility is pictured in Fig.\;\ref{fz7}(b), and in this case we would excite the detector for the firewall but not for the traditional hole. But to have this `other side' to the firewall we need to pass through a `wall' of  of Planck-scale physics. Since the strength of gravitational interactions increases with energy, we can expect that the largest interactions would be when the detector is crossing the region of Planck temperature, and so we cannot focus on the issue of detection of quanta with wavelength $\sim 10^6 l_p$  without asking if the theory allows us to pass through the Planck temperature region.  (In particular, it is hard to  imagine a physical model reproducing Fig.\;\ref{fz7}(c) which has smooth space on both sides of a Planck energy region.)
As we note below, the fuzzball microstates of string theory do {\it not} allow us to pass through the Planck temperature region; spacetime ends there in a stringy mess. Further, as we will note in part (iii) below, interaction with the Planck-scale degrees of freedom is not what precludes the kind of complementarity that we find with fuzzballs; instead, it is this interaction which transfers information to the collective modes of the fuzzball and leads to a complementary description.

(c) In Fig.\;\ref{fz7}(c) we depict the situation with fuzzballs. The incoming quanta cannot pass through the fuzzball surface, and so they transfer their energy to excitations of the fuzzball. The fuzzball details and the radiation it emits are parts of the same structure: the radiation is the  small time dependent part of the gravitational solution away from $r\approx 2M$. The response of the Planck-scale degrees of freedom in encoded in the response (\ref{evolve}) of the fuzzball, and this effect is expected to dominate over interactions with the radiation tail.

One thing is important to note about this interaction. Let the infalling observer be made of degrees of freedom that evolve slower than the Planck scale. Then the observer does not evolve significantly between the time that its coupling to the radiation becomes significant and the time it reaches the fuzzball boundary. Thus it is not clear what `burning' means in this context. The correct question to focus on is not the evolution of the infalling observer,  but rather the evolution (\ref{evolve}) of the fuzzball degrees of freedom that the observer impacts.

\b

\noindent{ {\bf (iii) The possibility of complementarity}}

Finally, let us address the issue of complementarity. The AMPS paper claims that their argument applies to the proposal of fuzzball complementarity. We can paraphrase this as claiming that, even for the simplest of hard-impact processes involving high-energy quanta, if an infalling quantum `burns up' at the Planck-temperature surface of the fuzzball, then there cannot be any other approximate complementary description involving free infall. This desire to avoid any interaction is suggested by the traditional proposal of complementarity. But as we have seen in Section \ref{trad}, there are difficulties with traditional complementarity, and this is not the kind of complementarity that we have proposed. 

Consider first our toy example of AdS/CFT duality. A graviton falling onto a D1D5 brane bound state {\it did } interact strongly on reaching these branes and broke up into a pair of excitations. Yet there was a complementary description where it passed smoothly into an AdS region. Similarly, $E\gg kT$ gravitons falling onto the fuzzball surface {\it do} interact strongly with the surface and excite the collective dynamics of the fuzzball degrees of freedom; it is this collective dynamics (\ref{evolve}) which will have a dual representation where to a first approximation the graviton will appear to fall through a horizon.

The moral we draw is that it is incorrect to conclude that complementarity would be impossible if an object encountered strong interactions near the horizon.  The situation is quite the opposite: we need {\it strong} interactions near the horizon to absorb the energy of the infalling quantum into the black hole's degrees of freedom to get `fuzzball complementarity'. If the absorption leads to an approximately faithful map of the infalling quantum's Hilbert space into a subspace of the black hole degrees of freedom, then we  have the possibility of a complementary description of the infall.

\section{Discussion}\label{secsix}

In this paper we have done two things: we summarized how complementarity is conjectured to work with fuzzballs, and we noted how the AMPS argument fails to address the underlying physics in this conjecture. In the discussion below we will put these two parts together, to see more directly where the AMPS argument goes wrong. In short, we will see that complementarity is a story of {\it two} descriptions of the physics, while AMPS try to have elements of both descriptions in the {\it same} setting. 

For the discussion below, it is helpful to summarize one version of the AMPS argument as follows:

\b

(1) Suppose the infalling observer sees nothing around $r=2M$  in some
description.

(2) Then in this description we have a smooth patch of spacetime around
the horizon.
 
(3) Evolution of vacuum modes in this smooth patch will lead to an entangled Hawking pair, and this
will lead to the information problem.

\b

The problem with this argument is that the description in which we have (1) (i.e. smooth spacetime) is valid only as an approximation for describing the physics of hard-impact $E\gg kT$ infalling quanta for short times (order $\sim M$). One cannot use the effective smooth spacetime used in this approximation to describe the entanglement of Hawking pairs over the much longer Hawking evaporation time (order $\sim M^2$); in particular one cannot  relate the effective description of (1) to the information problem which needs us to talk about the details of $\sim (M/m_p)^2$ Hawking pairs.

 Let us now see in more detail how things actually work:
 
 \b
 
 (a) The microstates of the black hole are fuzzballs, which means that the gravitational solution ends just outside $r=2M$ when the compact directions pinch off; the structure at this location is a quantum mess of KK monopoles, strings, fluxes, etc. (i.e. the set of allowed sources in string theory). 
 
 (b) $E\sim kT$ radiation is emitted from these sources, carrying the information of the microstate. A simple model to keep in mind is the computation of \cite{radiation}, where ergoregions near the fuzzball surface emit quanta by ergoregion emission. From this computation we learn that there is no sharp separation between the radiation and the fuzzball: the gravitational field in the ergoregion is unstable and radiates gravitons. If we follow these emitted gravitons back to their source, then we find more and more nonlinear gravitational physics, culminating in the `cap' where the fuzzball solution ends in KK monopoles etc. Thus whenever we ask if we interact with emitted quanta, we might as well go all the way and ask if we interact with the fully nonlinear `cap'. 
 
 (c) We recalled the toy example of AdS/CFT, which has similarities and differences with the black hole case. For now we look at the similarities.  Suppose we have a bound state of $N$ D1 and $N$ D5 branes. The infalling quanta of Fig.\;\ref{fz2p}(a) impacts this collection of branes and transfers its energy into excitations of the branes. Similarly, a quantum falling onto the fuzzball transfers its energy to the string theoretic sources (KK monopoles etc) on the fuzzball surface.  In (b) we had noted that the radiation from the fuzzball was just the tail end of the full nonlinear KK monopole `cap', so interactions with radiation near the horizon are included in this description.

(d) But in the AdS/CFT case, there is a second description, that of Fig.\;\ref{fz2p}(b), where the infalling quantum sails smoothly through into an AdS region. In this description we do not see the D1 and D5 branes as something that can be `hit'.  In the analogous case of fuzzballs, an infalling object does not see the nonlinear KK monopoles etc. near $r=2M$,  but instead sails through smoothly. In particular, it does not see the `tail end' of the nonlinear structure -- the radiation from the fuzzball -- as high energy quanta that can be `hit'. 

(e) To understand how it is possible for the infalling observer to sail through smoothly, consider first the AdS/CFT example. If the D1,D5 branes  were `inert'; i.e., they did not shift their internal state when the infalling object approached, then there would not be any description where the object `sailed through'. But in fact the D1D5 brane bound state has a {\it vast} space of internal excitations, and this changes the situation: the approach of the infalling object creates excitations in this vast space of possibilities, and the dynamics of these excitations is the dominant physics of the combined branes+object system. It is this dynamics that is described by the smooth infall into AdS space.\footnote{One often thinks of AdS/CFT duality as saying that the gravity variables in AdS can be re-expressed in terms of gauge theory variables on the boundary of AdS. But the origin of this duality is in the context of absorption by D-branes and black holes, and in that context the natural process to consider is the infall of quanta from infinity onto the branes. The largeness of $N$, the number of branes, leads to the excitation spectrum of the branes as being very dense, and the effective AdS description emerges.}

(f) The fuzzball has a similarly large phase space of deformations, since the number of fuzzball solutions is $Exp[S_{bek}]$. Now we see the basic element missing from the analysis of AMPS. They ask for the dynamics of the infalling object (what it measures etc.) but they ignore the fact that the much more important dynamics is the change of the state of the {\it fuzzball}: $\sum_i C_i |F_i\rangle\r \sum_j C'_j F'_j$.  This latter dynamics is so dominant that one must consider the infalling object and fuzzball as one unified system and then analyze the dynamics. When infalling quanta with energy $E\gg kT$ fall freely onto the fuzzball from far away, the conjecture is that the resulting dynamics has an approximate description valid for short times (order $\sim M$) that mimics infall through a smooth horizon \cite{plumpre,plumberg,otherfcrefs}; this is analogous to how in the  AdS/CFT case the object falls through smoothly into an AdS space. 

(g) The approximate nature of the `smooth infall' description is important. Since this description is valid only over a time of order $\sim M$, we cannot use this patch of smooth space to argue that entangled Hawking pairs will be created and will escape to large distances from the black hole. There is hardly time to create one pair in such a region. We cannot join together many such patches to argue that we have created many entangled pairs, since the description is only valid for short times and does not accurately track $E\sim kT$ physics.    The existence of many entangled pairs would have led to the information problem as discussed in Section \ref{basics}(b),(c); this problem does not arise here since we cannot study the creation of a large set of such pairs in our approximate `smooth infall' description.

\b

To summarize, the error in the AMPS argument can be seen by considering the infall of an observer into a stack of branes. These branes are in a particular  internal state, which can be probed by patient low energy scattering experiments from infinity. But the infalling observer reports none of this structure as he approaches the branes; he feels as if it is falling through empty AdS space. This `magical disappearance' of the branes can be traced to the fact that the branes have a vast set of internal states, and the dominant effect of the approach of the observer is to alter the internal state of the branes. Thus the dynamics of the 
branes+observer system is governed by the evolution of these newly created excitations, and not by the observer scattering off a fixed state of the branes. 

We can now see the fundamental role that the fuzzball construction plays in resolving the puzzles with black holes. If we have the traditional Penrose diagram of the hole, with vacuum at the horizon, then we get the creation of entangled pairs, and we cannot evade the Hawking information loss problem \cite{hawking,cern}. But in string theory we find that there is  very nontrivial structure at the horizon: the KK monopoles etc at the fuzzball surface carry `real' degrees of freedom that radiate unitarily
like a normal body. This resolves the information paradox. But we can ask a different question: what happens when we consider hard impacts of high energy ($E\gg kT$) quanta on the fuzzball surface? In this case the physics is analogous to what we find in AdS/CFT: the KK monopole and other string theoretic degrees of freedom on the fuzzball surface act like the branes in the D1D5 system. The infalling observer reports nothing special as it approaches these objects, since the dominant dynamics is that of {\it exciting} the fuzzball degrees of freedom, not the response of the observer. AMPS implicitly assume that they are falling towards a radiating surface that is {\it inert} to such excitations, and thus miss the physics of free infall which is common to the fuzzball and the AdS/CFT cases.

In the context of the argument (1)-(3) listed at the start of this section, we see that the implication (1) $\rightarrow$ (2) is misleading. It is not that we don't have structure at the location of the branes; rather, the infalling observer does not report such structure. The patch of smooth spacetime in (2) is an effective description of the $\sum_i C_i |F_i\rangle\r \sum_j C'_j F'_j$ dynamics which describes the excitations of the impacted fuzzballs; it is not the actual gravitational solution at the horizon. The implication (2) $\rightarrow$ (3) does not work since this effective description cannot be applied to a region larger than $\sim M$ which would be needed to create a large number of entangled pair. In general there are {\it two} descriptions involved: (i) the actual microscopic fuzzball which carries all information of the state and radiates unitarily, and  (ii) the approximate short time description of collective modes, that mimics free infall. AMPS do not differentiate carefully these two descriptions, and that leads them to claim an apparent contradiction with fuzzball complementarity. 

In conclusion, the AMPS argument does not apply to the process by which complementarity is conjectured to arise in the fuzzball picture. But it is a very interesting argument to consider, since it brings out clearly the various important physical principles involved in the quantum dynamics of black holes. 

\section*{Acknowledgements}

This work was supported in part by DOE grant DE-FG02-91ER-40690. We thank the authors of \cite{amps} as well as Iosif Bena, Borun Chowdhury, Stefano Giusto, Oleg Lunin, Lenny Susskind and  Nick Warner for discussions.  In particular we are grateful to Don Marolf for patiently explaining to us the nature of the AMPS argument.

\begin{appendix}

\section[Appendices]{Timescale for detection} \label{detect}

Here we note that a detector needs a proper time  $\Delta \tau\gtrsim \lambda$ to detect a quantum of wavelength $\lambda$. Since this argument is well known, we will describe it for the simple case of a detector at rest in the Minkowski vacuum; the extension to other situations is straightforward.

We assume for simplicity that the metric is time independent in our choice of coordinates. The field operator can be expanded as
\be
\hat\Phi=\sum_k [{1\over \sqrt{2\omega_k}} e^{i(kx-\omega_k t)}\hat a_k
+{1\over \sqrt{2\omega_k}}e^{-i(kx-\omega_k t)}\hat a_k^\dagger], ~~~~[\hat a_k, \hat a_{k'}^\dagger ] =\delta _{k, k'}
\ee
 We take the detector to be a harmonic oscillator
\be
\hat\Psi={1\over \sqrt{2\Omega}} e^{-i\Omega \tau}\hat A + {1\over \sqrt{2\Omega}} e^{i\Omega \tau}\hat A^\dagger, ~~~[\hat A, \hat A^\dagger]=1
\ee
The interaction along the worldline is given by $\int d\tau \, \hat H_{int} (\tau)$ where
\be
\hat H_{int}(\tau)=q \, h(\tau) \, \hat\Phi\bigl (t(\tau), x(\tau)\bigr )\hat \Psi(\tau)
\ee
Here $q$ is a coupling constant and $0\le h(\tau)\le 1$ is a function that allows us to switch on and switch off the detector.

We start at $\tau\r -\infty$ with the detector in the ground state: $\hat A|0\rangle_A=0$. Let us also take the spacetime to be empty of quanta: $\hat a_k |0\rangle_a=0$. We take first order perturbation theory in  $q$. The amplitude to reach the state $|1\rangle_A|1\rangle_{k} \equiv \hat A^\dagger \hat a_k^\dagger |0\rangle_A|0\rangle_a$ is
\be
{\cal A}=-iq\int_{-\infty}^\infty d\tau {1\over \sqrt{2\omega_k}}{1\over \sqrt{2\Omega}}h(\tau)e^{i\Omega \tau}e^{-ikx(\tau)+i\omega_k t(\tau)}
\ee
We take
\be
h(\tau)=e^{- ({\tau\over \Delta \tau})^2}
\ee
which corresponds to making a measurement over an interval $\sim \Delta\tau$. 
We also let the detector trajectory to describe a detector at rest at $x=0$, which gives $x(\tau)=0, t(\tau)=\tau$ for all $\tau$. This gives 
\be
{\cal A}=-i q\int _{-\infty}^\infty d\tau {1\over \sqrt{2\omega_k}}{1\over \sqrt{2\Omega}}e^{- ({\tau\over \Delta \tau})^2}e^{i(\Omega+\omega_k)\tau}=-iq{1\over \sqrt{2\omega_k}}{1\over \sqrt{2\Omega}}\Delta\tau\sqrt{\pi}e^{-{1\over 4}(\Delta \tau)^2(\Omega+\omega_k)^2}
\ee
Keeping the detector on for all time is equivalent to taking $\Delta\tau\r \infty$, in which case we get ${\cal A}=0$. So the detector does not get excited, which is expected since we started with empty Minkowski space. 

But now consider a situation where the detector is switched on and off in a comparatively short interval, as would need to be the case if one was trying to detect a Hawking quantum by an infalling detector before the detector hit the black hole surface. For detection times shorter than the wavelengths we want to measure
\be
\Delta\tau\lesssim {1\over (\Omega+\omega_k)}
\ee
we get
\be
{\cal A}\sim -iq{1\over \sqrt{2\omega_k}}{1\over \sqrt{2\Omega}}\Delta\tau\sqrt{\pi}\ne 0
\ee
so we pick up vacuum fluctuations in the detector. 

To summarize, suppose we make a detector with frequency $\Omega$ to pick up quanta of wavelength $\lambda\sim \Omega^{-1}$. Then the effect of vacuum fluctuations will be comparable to the  effect of  `real quanta' if
\be
\Delta\tau \lesssim {1\over (\Omega+\omega_k)} < {1\over \Omega}\sim \lambda
\ee

\refstepcounter{section}
\section*{\thesection \quad Wavelength of Hawking quanta} \label{wavelength}

Consider the Schwarzschild black hole
\be
ds^2=-(1-{2M\over r})dt^{2}+{dr^2\over 1-{2M\over r}}+r^{2}{d\Omega_2^{2}}
\label{oneq}
\ee
The temperature is ${1\over 8\pi M}$, so the wavelength of Hawking quanta at infinity is $\lambda_\infty\sim M$. The wavelength of such a quantum at any position $r$ is
\be
\lambda\sim (-g_{tt})^\h\lambda_\infty\sim M(1-{2M\over r})^\h
\ee
Near the horizon $(r-2M)\ll 2M$ we can use Rindler coordinates
\be
t_R={t\over 4M}, ~~r_R=\sqrt{8M(r-2M)}
\label{sixt}
\ee
This gives  the metric in the time and radial directions
\be
ds^2\approx -r_R^2 dt_R^2+dr_R^2
\label{fift}
\ee
From now on we restrict attention to just these directions. 
In this near-horizon region we have for the wavelength of radiated quanta
\be
\lambda\sim M^\h (r-2M)^\h\sim r_R
\ee
From (\ref{fift}) we see that the distance from the horizon measured on a constant $t_R$ slice is $d=r_R$. Thus if a black hole emits radiation at the Hawking temperature, then the wavelength of these quanta at a distance $d$ from the horizon is 
\be
\lambda\sim d
\ee
This is the wavelength measured along a slice of constant Schwarzschild time $t$. If this quantum is encountered by an infalling detector, then the effective wavelength will be Lorentz contracted. Let the proper velocity of the detector  
in a local Lorentz frame oriented along the Schwarzschild $t, r$ directions, be  
\be
U^{\hat t}=\cosh\alpha, ~~~U^{\hat r}=-\sinh\alpha
\label{eqalpha}
\ee
The momentum vector of an outgoing massless  quantum in the local Lorentz frame is
\be
(p^{\hat t}, p^{\hat r})\sim ({1\over \lambda}, {1\over \lambda})
\ee
The energy of the quantum as measured by the detector is then
\be
E=-p_\mu U^\mu\sim {1\over \lambda}(\cosh\alpha+\sinh\alpha)={1\over \lambda} e^\alpha
\ee
and the effective wavelength that is seen by the infalling detector is then
\be
\lambda_{eff}\sim \lambda e^{-\alpha}\sim d e^{-\alpha}
\label{sevent}
\ee
where as above, $d$ is the distance measured from the horizon in the Schwarzschild frame along a $t=const$ slice.

\refstepcounter{section}
\section*{\thesection \quad Proper time along infalling geodesic} \label{time}

We wish to ask how much proper time $\Delta \tau$ elapses along a geodesic between the time it is at a distance $d$ from $r=2M$ and the time it hits the black hole surface at $r=2M$. Since we are working near the horizon, we use the Rindler coordinates (\ref{sixt}). The Kruskal-type coordinates appropriates to a freely falling observer are given locally by taking the Minkowski coordinates related to $t_R, r_R$ by
\be
t_M=r_R \sinh t_R, ~~~x_M=r_R \cosh t_R
\ee

 \begin{figure}[t]
\begin{center}
\includegraphics[scale=.65]{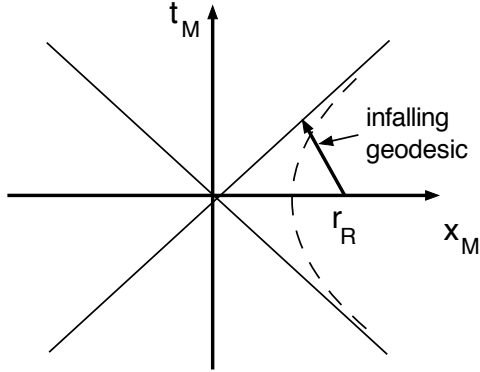}
\end{center}
%
%
\caption{The Rindler coordinates near the horizon, and the corresponding Minkowski coordinates. The infalling geodesic starts at $r_R=d, t_R=0$ and ends at the Rindler horizon.}
\label{fz3}       
\end{figure}

Fig.\;\ref{fz3} shows the geodesic that we follow. This geodesic is a straight line in the local Minkowski coordinates
\be
t_M=\cosh\alpha ~\tau, ~~~x_M=-\sinh\alpha~\tau+d
\ee
We have taken the geodesic to start with $\tau=0$ at position $r_R=d$ and time $t_R=0$.
Here $\alpha$ is a constant the gives the velocity of infall; note that it is the same $\alpha$ as the one that appears in (\ref{eqalpha}).  The geodesic crosses the horizon $t_M=r_M$ at proper time $\tau_f$ with
\be
\cosh\alpha ~\tau_f =-\sinh\alpha ~ \tau_f + d, ~~~\Rightarrow~~~\tau_f=d e^{-\alpha}
\ee
 Thus if an observer on an infalling trajectory tries to detect a quantum at distance $d$ from the horizon, then the time he has available to make the detection is
\be
\Delta\tau_{available}< \tau_f=d e^{-\alpha}
\ee

\end{appendix}

\end{document}